# Bulk superconductivity in $La_2O_2M_4S_6$-type layered oxychalcogenide $La_2O_2Bi_3Ag_{0.6}Sn_{0.4}S_{5.7}Se_{0.3}$

Rajveer Jha[1], Yosuke Goto[1], Tatsuma D. Matsuda[1], Yuji Aoki[1], Masanori Nagao[2], Isao Tanaka[2], and Yoshikazu Mizuguchi[1*]

[1]*Department of Physics, Tokyo Metropolitan University, 1-1 Minami-Osawa, Hachioji, Tokyo 192-0397, Japan.*

[2]*University of Yamanashi, 7-32, Miyamae, Kofu, Yamanashi 400-8511, Japan.*

Corresponding author: Yoshikazu Mizuguchi (mizugu@tmu.ac.jp)

Abstract:

Recently, we reported the observation of superconductivity at ~0.5 K in a $La_2O_2M_4S_6$-type (M: metal) layered oxychalcogenide $La_2O_2Bi_3AgS_6$, which is a layered compound related to the $BiS_2$-based superconductor system but possesses a thicker $Bi_3AgS_6$-type conducting layer. In this study, we have developed the $La_2O_2Bi_3AgS_6$-type materials by element substitutions to increase the transition temperature ($T_c$) and to induce bulk nature of superconductivity. A resistivity anomaly observed at 180 K in $La_2O_2Bi_3AgS_6$ was systematically suppressed by Sn substitution for the Ag site. By the Sn substitution, $T_c$ increased, and the shielding volume fraction estimated from magnetization measurements also increased. The highest $T_c$ (= 2.3 K) and the highest shielding volume fraction (~20%) was observed for $La_2O_2Bi_3Ag_{0.6}Sn_{0.4}S_6$. The superconducting properties were further improved by Se substitutions for the S site. By the combinational substitutions of Sn and Se, bulk-superconducting phase of $La_2O_2Bi_3Ag_{0.6}Sn_{0.4}S_{5.7}Se_{0.3}$ with a $T_c$ of 3.0 K ($T_c^{onset}$ = 3.6 K) was obtained.

**Introduction**

Layered superconductors have been extensively studied due to observations of unconventional superconductivity and high transition temperature ($T_c$) [1,2]. In addition, the great flexibility of constituent elements and layered (stacking) structure, the studies on layered superconductor system can be widely developed. The recent discovery of BiS$_2$-based superconductors has also created remarkable attention in the superconductivity community: the typical materials are Bi$_4$O$_4$S$_3$, REO$_{1-x}$F$_x$BiS$_2$ (RE = La, Ce, Pr, Nd, Yb), and Sr$_{1-x}$RE$_x$FBiS$_2$ [3-14]. For instance, the crystal structure of the typical system LaOBiS$_2$ is composed of alternate stacks of a La$_2$O$_2$ blocking layer and two BiS$_2$ layers. Since the parent phase (LaOBiS2) is an insulator with a band gap [15], electron doping is needed to induce metallic and superconducting characteristics. Furthermore, local in-plane structure should be optimized to induce bulk superconductivity in the BiS$_2$ layer [16]. Thus, the $T_c$ of BiS$_2$-based is sensitive to physical pressure [4, 17-21] and chemical pressure [22-24] effects and reaches 11 K in LaO$_{0.5}$F$_{0.5}$BiS$_2$. Therefore, an increase in the highest record of $T_c$ in the BiS$_2$-based superconductor family can be expected by further material development and tuning of structural and electrical properties [14].

Very recently, we reported on the superconductivity in oxychalcogenide La$_2$O$_2$Bi$_3$AgS$_6$ with $T_c$ of 0.5 K [25]. The crystal structure of La$_2$O$_2$Bi$_3$AgS$_6$ is similar to the typical BiS$_2$-based superconductor LaOBiS$_2$, but the conducting layer thickness is thicker than that of LaOBiS$_2$. The crystal structure of the system can be described as La$_2$O$_2$M$_4$S$_6$, in which the M site can be Pb, Bi, and Ag [26-28]. There are two different metal sites [M1 and M2 displayed in Fig. 1(c)] in the M$_4$S$_6$-type conducting layer. For the outer layers with the M1 site, Bi selectively occupies, and the layers can be regarded as the BiS$_2$-type layers. For the inner layer with the M2 site, NaCl-type (Bi,Ag)S$_2$ layers are inserted in between the BiS$_2$-type layers.

Interestingly, an anomalous transport property was observed in the temperature dependence of resistivity in La$_2$O$_2$Bi$_3$AgS$_6$ [25], which is similar to the charge-density-wave (CDW) transition in EuFBiS$_2$ [29]. In this study, we have investigated the Sn substitution effect for the Ag site in La$_2$O$_2$Bi$_3$Ag$_{1-x}$Sn$_x$S$_6$ and found that the resistivity anomaly was suppressed by Sn substitutions. $T_c$ reached 2.3 K for $x$ = 0.4 (La$_2$O$_2$Bi$_3$Ag$_{0.6}$Sn$_{0.4}$S$_6$). Furthermore, by Se substitutions for the S site in La$_2$O$_2$Bi$_3$Ag$_{0.6}$Sn$_{0.4}$S$_{5.7}$Se$_{0.3}$, the $T_c$ further increased to 3.5 K, and bulk nature of superconductivity was confirmed from magnetic shielding volume fraction.

**Results**

**Sn substitution effect on structural and physical properties in $La_2O_2Bi_3Ag_{1-x}Sn_xS_6$**

Figure 1a displays the room temperature XRD pattern for $La_2O_2Bi_3Ag_{1-x}Sn_xS_6$ ($x = 0$–0.5). All the $La_2O_2Bi_3Ag_{1-x}Sn_xS_6$ samples are crystallized in the tetragonal structure with the space group of *P*4/*nmm*. An impurity phase of $La_2Sn_2O_7$ was observed for $x = 0.2$–0.5. Fig. 1b shows the shift in the 103 peak position, which shifts towards the low angle side for $x = 0.1$. As the Sn concentration increases from $x = 0.2$ to 0.5, the 103 peak shifts towards the higher angle side. The schematic image of the crystal structure of $La_2O_2M_4S_6$ is shown in Fig. 1c. For the $La_2O_2Bi_3Ag_{1-x}Sn_xS_6$ phase, Bi selectively occupies the M1 site, and Sn is expected to substituted for Ag at the M2 site, which was qualitatively confirmed by Rietveld refinements. The evolutions of the lattice parameters by the Sn substitution are shown in Figs.1d and 1e. The lattice parameters are $a = 4.061(1)$ Å and $c = 19.445(1)$ Å for $x = 0$ and $a = 4.0648(1)$ Å and $c = 19.48(1)$ Å for $x = 0.1$. The lattice parameter $c$ tends to decrease by Sn substitution for $x = 0$–0.2, and then, it continuously increases with increasing $x$ for $x = 0.2$-0.5. The lattice parameter $a$ increases for the $x = 0.1$, but it tends to decrease with increasing $x$ for higher $x$, but the change is small. On the basis of the lattice parameter evolutions, the Sn substitution does not largely affect the lattice volume, which may be due to smaller ionic radius of $Sn^{2+}$ (93 pm) than that of $Ag^+$ (113 pm). Since the structure (bonding) in the LaO layer is not flexible as compared to the conducting layer, the unit cell volume is mainly affected by the blocking layer in the $BiS_2$-based systems. The actual ratio of the metals in the conducting layer (Bi, Ag, and Sn) in the $La_2O_2Bi_3Ag_{1-x}Sn_xS_6$ ($x = 0$–0.5) samples were examined by EDX. The nominal and analyzed values of the $x$ are plotted in the Fig. 1f. Although there is slight deviation from the nominal values, the analyzed values of $x_{EDX}$ linearly increases with increasing nominal $x$. We found that Bi is slightly excess for $x = 0.1$–0.3. Therefore, $y$ parameter with a formula of $La_2O_2Bi_{3+y}(Ag_{1-x}Sn_x)_{1-y}S_6$ was introduced to analyze the Bi concentration. The analyzed values of $y_{EDX}$ are plotted in Fig. 1g. $y_{EDX}$ is higher for $x = 0.1$–0.3 but almost zero for $x = 0, 0.4$, and 0.5. Therefore, we consider that the Bi excess can be ignored in the discussion on the Sn substitution (and Se substitution) effect, and we use the formula $La_2O_2Bi_3Ag_{1-x}Sn_xS_6$ in this paper

Figures 2a-c show the temperature dependences of magnetic susceptibility ($4\pi\chi$-$T$) under an applied magnetic field of 10 Oe for $La_2O_2Bi_3Ag_{1-x}Sn_xS_6$ ($x = 0.3$–0.5). The diamagnetic signals in the $4\pi\chi$ curve were observed below 2.2, 2.8, and 2.6 K for $x = 0.3$–0.5 respectively. A large

diamagnetic signal was observed below 2.8 K in the ZFC curve for $x = 0.4$. The shielding value fractions estimated from $4\pi\chi$ (ZFC) at 1.9 K is nearly 20% [See Fig. 2(d).] while it is still not saturated. From the susceptibility results, we consider that Sn substitution is effective to improve the superconducting properties of $La_2O_2Bi_3AgS_6$ but not sufficient to induce bulk superconductivity.

Figure 3 shows the temperature dependences of electrical resistivity from 300 to 0.1 K for $La_2O_2Bi_3Ag_{1-x}Sn_xS_6$ ($x = 0$–0.5). The electrical resistivity at 300 K decreases with increasing Sn concentration up to $x = 0.3$ and increases again for $x = 0.4$ and 0.5. The normal-state resistivity of the Sn-doped samples changes remarkably. For example, the pure sample ($x = 0$) shows a linear decrease in resistivity on cooling below the anomaly temperature $T^* = 180$ K. A similar behavior was observed up to $x = 0.2$. The resistivity anomaly at $T^*$ appears for $x \leq 0.2$, and the $T^*$ shifts towards the lower temperature side with increasing $x$. In contrast, the normal-state $\rho(T)$ for $x = 0.3$–0.5 shows an upturn below ~50 K. The anomaly disappears for $x \geq 0.3$. Figure 3g shows the zoomed view of the Figs. 3a-3f near the superconducting transition. The $T_c$ clearly increases with increasing Sn concentration in $La_2O_2Bi_3Ag_{1-x}Sn_xS_6$. The highest $T_c$ was achieved for $x = 0.4$, and $T_c$ decreases for a higher substitution with $x = 0.5$.

The room-temperature Seebeck coefficient ($S$) for $La_2O_2Bi_3Ag_{1-x}Sn_xS_6$ ($x = 0$–0.5) are shown in Fig. 4. The Seebeck coefficient is a good scale for the carrier concentration in $BiS_2$-based compounds [30]. We observed a slight change in $S$ by Sn substitution. The $S$ in $x = 0.2$–0.4 are almost the same, but that for $x = 0$ and 0.5 are slightly large. This suggests that the carrier concentrations for $x = 0.2$–0.4 are higher than those for $x = 0$ and 0.5. This seems to be related to the evolution of $T_c$. However, the large change in $T_c$ from 0.6 to 2.3 K between $x = 0.1$ and 0.4 cannot be simply understood by the carrier concentration only.

Figure 5 shows the superconductivity phase diagram of $La_2O_2Bi_3Ag_{1-x}Sn_xS_6$, which shows the interplay between the resistivity anomaly temperature ($T^*$) and the superconducting transition temperature ($T_c^{zero}$). The $T^*$ is suppressed by the Sn substitution, and it disappears at $x = 0.3$. The $T_c$ gradually increases with increasing $x$ in $La_2O_2Bi_3Ag_{1-x}Sn_xS_6$. The highest $T_c^{zero} = 2.3$ K is achieved for $x = 0.4$. A lower $T_c^{zero} = 1.9$ K is observed for the highest (solubility-limit) Sn concentration of $x = 0.5$.

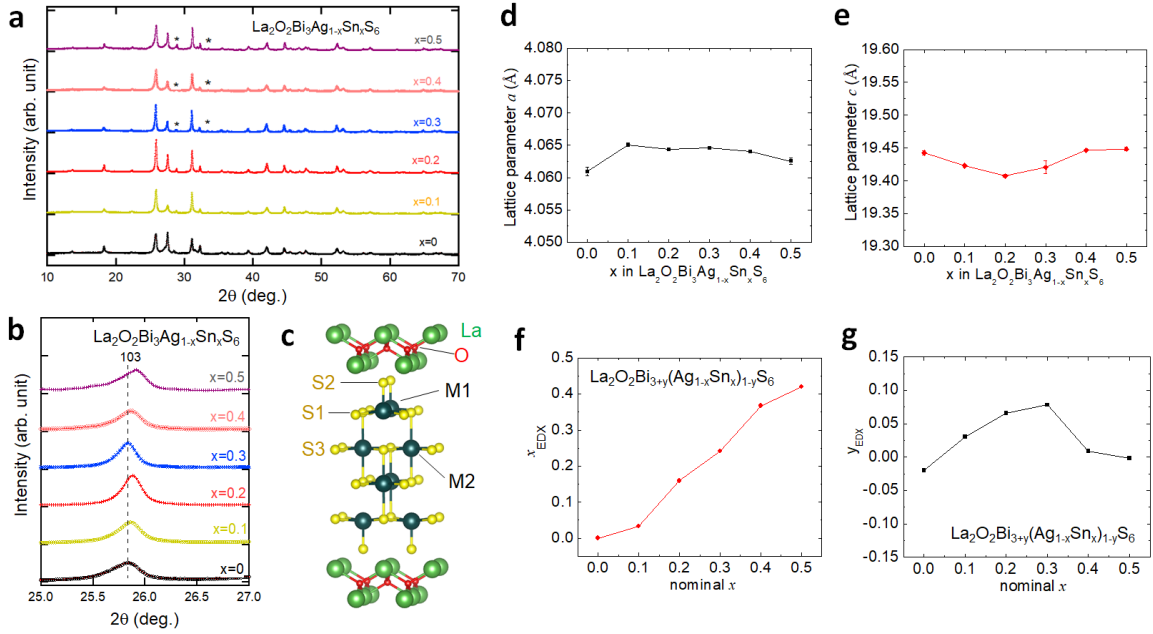

**Fig. 1. Crystal structure and composition analyses results for La$_2$O$_2$Bi$_3$Ag$_{1-x}$Sn$_x$S$_6$ ($x$ = 0–0.5).** **a.** Room-temperature XRD pattern for $x$ = 0–0.5. The impurity peaks of La$_2$Sn$_2$O$_7$ are indicated by asterisks (*). **b.** XRD pattern near the 103 (Miller index) peak of the tetragonal phase of La$_2$O$_2$Bi$_3$AgS$_6$. **c.** A schematic image of the crystal structure of La$_2$O$_2$M$_4$S$_6$ (two M sites, M1 and M2, are occupied by Bi, Ag, and Sn in the present system). **d,e.** Lattice parameters of $a$ and $c$ obtained from Rietveld refinements. **f, g.** Nominal composition dependences of compositions ($x$ and $y$) analyzed by EDX, where $x$ and $y$ are defined as La$_2$O$_2$Bi$_{3+y}$(Ag$_{1-x}$Sn$_x$)$_{1-y}$S$_6$.

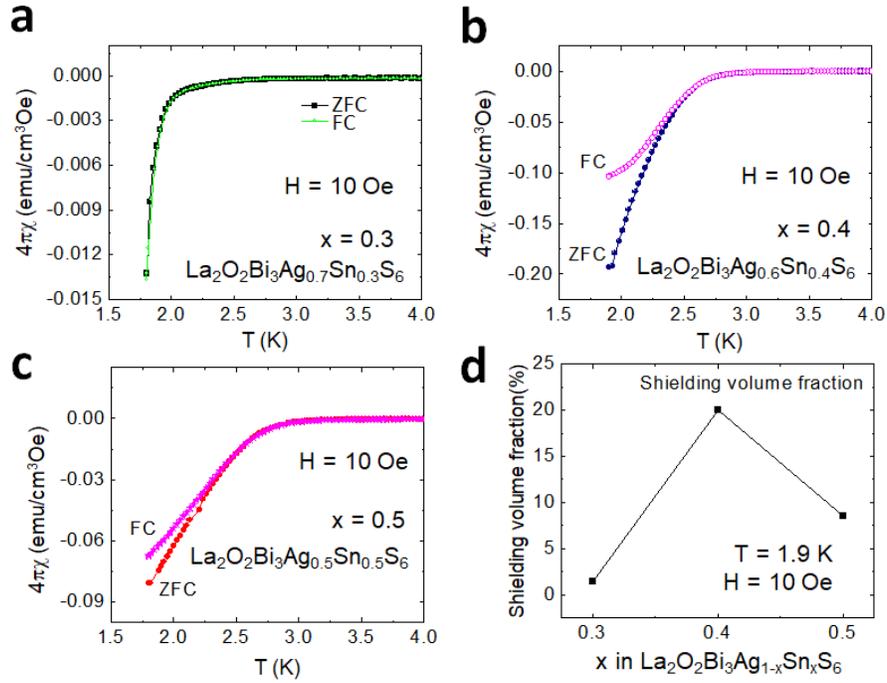

**Fig. 2. Superconducting properties examined from magnetic susceptibility for $La_2O_2Bi_3Ag_{1-x}Sn_xS_6$ ($x$ = 0.3–0.5). a–c.** Temperature ($T$) dependences of magnetic susceptibility ($4\pi\chi$) for $x$ = 0.3–0.5 measured in the ZFC and FC modes with an applied magnetic field of 10 Oe. **d.** Sn concentration dependence of the shielding volume fraction estimated using the ZFC data at 1.9 K.

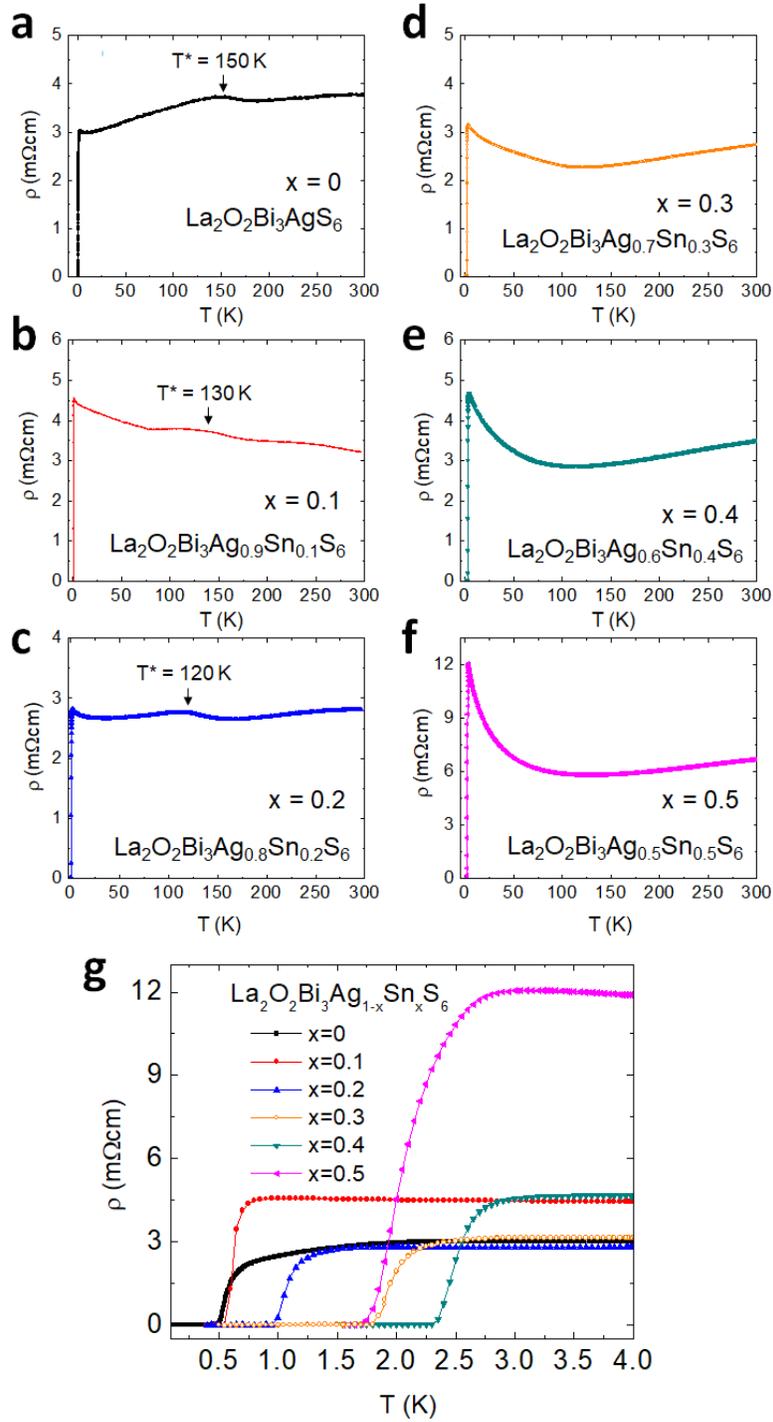

**Fig. 3. Electrical transport properties for La$_2$O$_2$Bi$_3$Ag$_{1-x}$Sn$_x$S$_6$ ($x$ = 0–0.5). a–f.** Temperature dependences of electrical resistivity from 300 to 0.1 K for $x$ = 0–0.5. The anomaly temperature in the $\rho(T)$ curves is indicated by $T^*$. **g.** The $\rho(T)$ curves in the temperature range of 0.1–4.0 K.

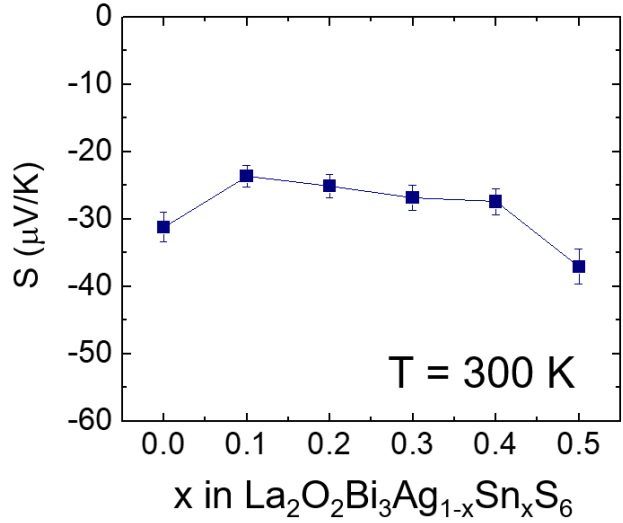

**Fig. 4. Seebeck coefficient for La$_2$O$_2$Bi$_3$Ag$_{1-x}$Sn$_x$S$_6$ ($x$ = 0–0.5).** The room-temperature Seebeck coefficient ($S$) is plotted as a function of nominal Sn concentration ($x$).

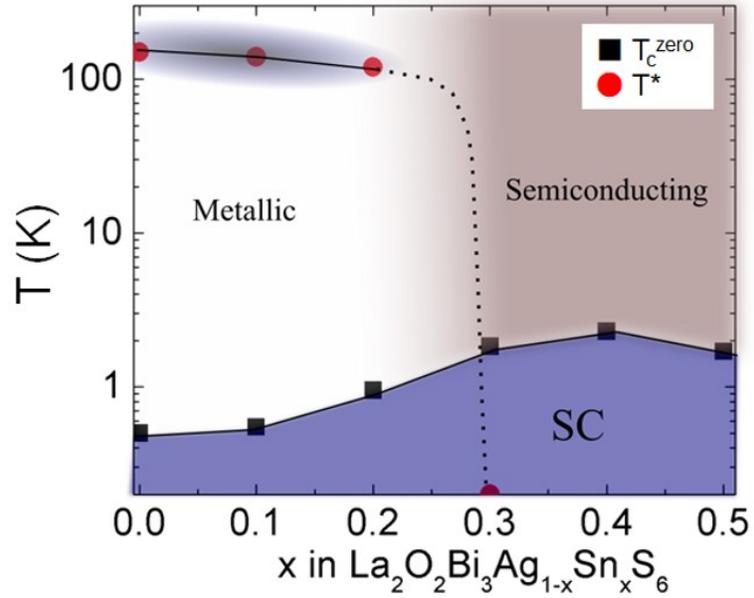

**Fig. 5. Superconductivity phase diagram for La$_2$O$_2$Bi$_3$Ag$_{1-x}$Sn$_x$S$_6$ ($x$ = 0–0.5).** The $x$ dependences of $T_c^{zero}$ and $T^*$ are plotted as a function of Sn concentration ($x$). SC denotes superconductivity.

**Superconducting properties of Se-doped $La_2O_2Bi_3Ag_{0.6}Sn_{0.4}S_{5.7}Se_{0.3}$**

As shown above, the Sn substitution improved the superconducting properties in $La_2O_2Bi_3Ag_{1-x}Sn_xS_6$, and the highest $T_c$ and shielding volume fraction were obtained for $x = 0.4$. In the $BiS_2$-based compounds, partial Se substitutions for the S site of superconducting $BiS_2$ layers have significantly improved the superconducting properties and the bulk characteristics of superconductivity. Therefore, we tried to substitute the S site of the $x = 0.4$ sample. The 5%-Se sample $La_2O_2Bi_3Ag_{0.6}Sn_{0.4}S_{5.7}Se_{0.3}$ was successfully synthesized, but samples with higher Se concentration contained selenide impurity phases. The solubility limit of Se for the S site is around 5%. The composition estimated from the EDX analyses for Bi, Ag, Sn, S, Se elements was $La_2O_2Bi_{3.09}Ag_{0.65}Sn_{0.26}S_{5.73}Se_{0.27}$. Since the obtained composition is close to the nominal formula, we call the sample with the nominal value below.

Figure 6 shows the XRD pattern and the Rietveld refinement result for $La_2O_2Bi_3Ag_{0.6}Sn_{0.4}S_{5.7}Se_{0.3}$. Although two peaks related to the $La_2Sn_2O_7$ impurity phase were observed, other peaks could be refined using the tetragonal ($P4/nmm$) model with a reliability factor $R_{wp}$ of 13.4%. In the refinement, Se was assumed to be substituted for the S1 site. The lattice parameters were $a = 4.0759(2)$ Å and $19.4824(11)$ Å, which are clearly larger than those of $La_2O_2Bi_3Ag_{1-x}Sn_xS_6$ due to the presence of Se.

Figure 7 displays the superconducting properties of $La_2O_2Bi_3Ag_{0.6}Sn_{0.4}S_{5.7}Se_{0.3}$. As shown in Fig. 7a, a large shielding volume fraction close to 100% was observed. From the resistivity measurements (Fig. 7b), zero resistivity was observed at 3.0 K, and the onset temperature ($T_c^{onset}$) was 3.5 K; we estimated the temperature where the resistivity becomes almost 90% of normal-state resistivity. Although superconductivity was observed, the $\rho(T)$ curve still shows a semiconducting-like localization at low temperatures. We have measured $\rho(T)$ under magnetic fields up to 9 T. The obtained $T_c^{onset}$ and $T_c^{zero}$ were plotted in Fig. 7d to evaluate the upper critical field $H_{c2}$ and the irreversible field $H_{irr}$. The $H_{c2}(0)$ was estimated as 2.15 T using the WHH model (Werthamer-Helfand-Hohenberg model) [31]. In addition, from rough estimation with a linear fitting of $H_{irr}$, $H_{irr}(0)$ was estimated as 1.0 T.

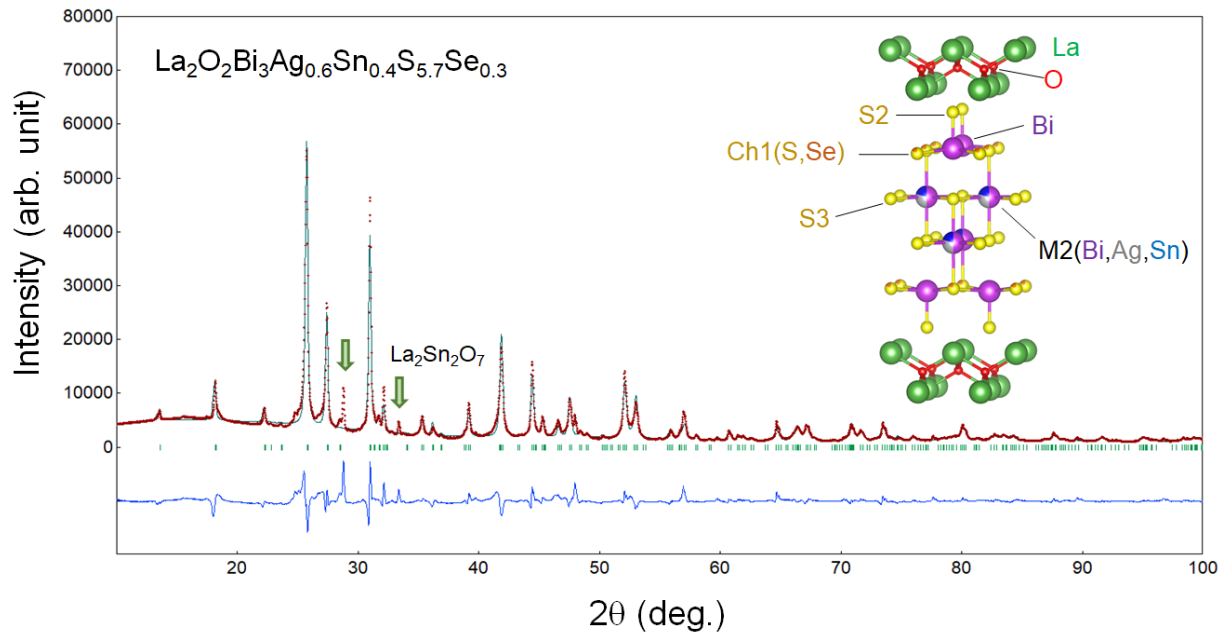

**Fig. 6. X-ray diffraction analysis for $La_2O_2Bi_3Ag_{0.6}Sn_{0.4}S_{5.7}Se_{0.3}$.** XRD pattern and the Rietveld refinement result are shown. The arrows indicate the peaks for the impurity phase $La_2Sn_2O_7$. The inset image shows the crystal structure depicted using the structural parameters obtained from the Rietveld refinement.

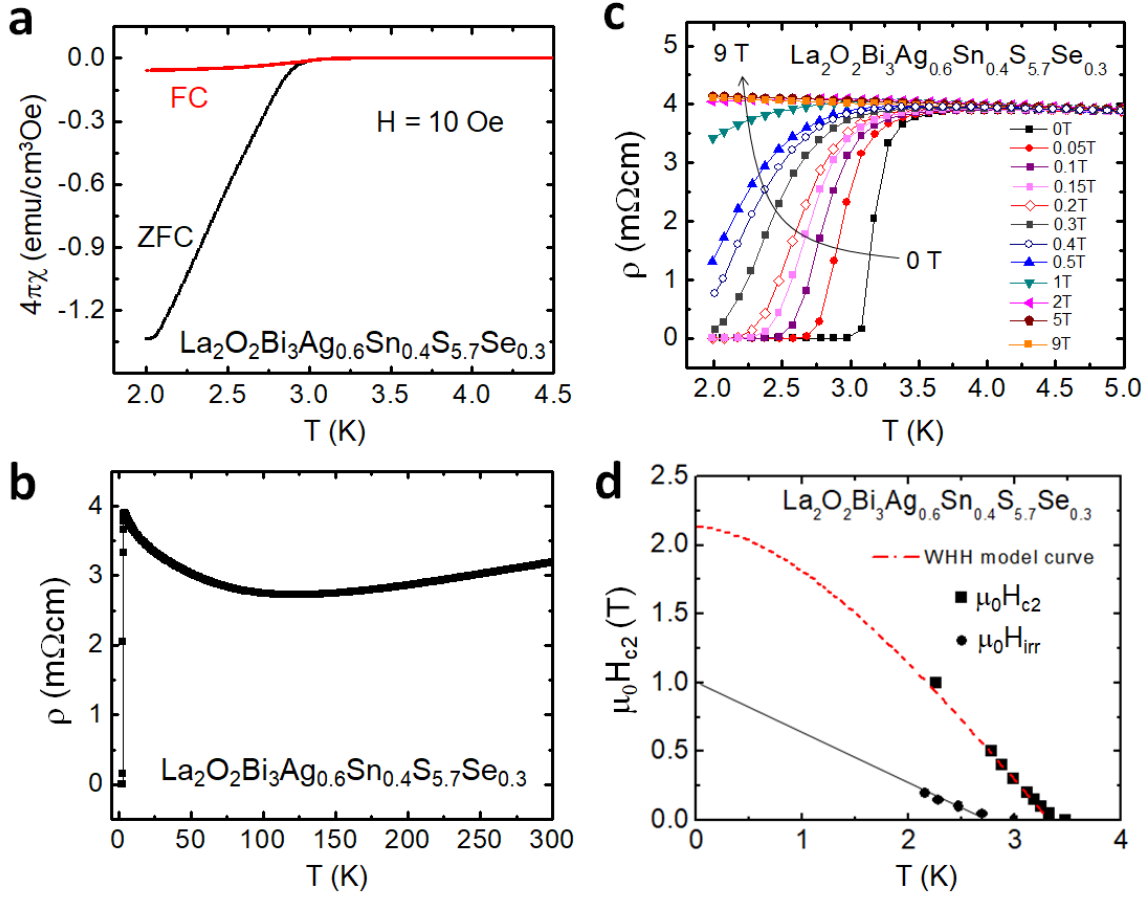

**Fig. 7. Superconducting properties for La$_2$O$_2$Bi$_3$Ag$_{0.6}$Sn$_{0.4}$S$_{5.7}$Se$_{0.3}$. a.** Temperature dependence of magnetic susceptibility. **b.** Temperature dependence of electrical resistivity [$\rho(T)$]. **c.** Low-temperature $\rho(T)$ under magnetic fields up to 9 T. **d.** Temperature-magnetic field phase diagram with the upper critical field ($H_{c2}$) and the irreversible field ($H_{irr}$).

## Discussion

### Suppression of resistivity anomaly by the Sn substitution

Here, we discuss the possible origin of the increase in $T_c$ by the Sn substitution. As revealed in the crystal structure part, the lattice parameters were not largely affected by the Sn substitution. Therefore, in-plane chemical pressure amplitude in the Bi-S superconducting plane, which has been revealed as the essential parameter for the emergence of superconductivity in BiS$_2$-based compounds [24], should not be significantly changed. Therefore, we consider that the in-plane chemical pressure effect is not the origin for the increase in $T_c$ by the Sn substitution. On carrier concentration, the absolute value of the Seebeck coefficient slightly decreases by Sn substitution for $x$ = 0.1–0.4, which can be corresponding to the slight increase in electron carriers by Sn substitution. However, the large increase in $T_c$ for $x$ = 0.4 may not be understood from the increase in carrier concentration only because the difference in carrier concentration between $x$ = 0.1 ($T_c$ = 0.6 K) and $x$ = 0.4 ($T_c$ = 2.3 K) is expected to be quite small. On the basis of these facts, we briefly mention about the possible relation to the CDW ordering and the possible scenario of the suppression of CDW by Sn substitution in this system. In the $\rho$-$T$ plots, an anomaly was observed in the La$_2$O$_2$Bi$_3$Ag$_{1-x}$Sn$_x$S$_6$ system. A similar feature in the normal-state resistivity has been observed in the EuFBiS$_2$ superconductor ($T_c$ = 0.3 K). The origin of the hump was proposed as a CDW transition [29]. We assume that the suppression of the CDW ordering is the origin for the increased $T_c$ in La$_2$O$_2$Bi$_3$Ag$_{1-x}$Sn$_x$S$_6$. In addition, the anomaly temperature $T^*$ was shifted to a lower temperature by Sn substitution, and the anomaly disappeared at $x$ = 0.3. At around $x$ = 0.3 and 0.4, $T_c$ is the maximum. These facts imply that $T_c$ increased by the suppression of $T^*$. Although we have no evidence for the CDW states in the La$_2$O$_2$Bi$_3$Ag$_{1-x}$Sn$_x$S$_6$ system and the suppression mechanism by the Sn substitution, introduction of randomness at the M2 site may be effective to suppress the charge ordering states.

**Bulk superconductivity in La$_2$O$_2$Bi$_3$Ag$_{0.6}$Sn$_{0.4}$S$_{5.7}$Se$_{0.3}$**

As shown in the Result part, a partial Se substitution for S induced bulk superconductivity in La$_2$O$_2$Bi$_3$Ag$_{0.6}$Sn$_{0.4}$S$_{5.7}$Se$_{0.3}$. Although the solubility limit is very low (5%), the lattice parameters clearly changed by the partial Se substitution, and the superconducting properties were significantly improved. Although we refined three models with different Se site (assuming the substitution at the S1, S2, or S3 sites), we could not find the site selectivity of doped Se. However, we expect that the doped Se occupies the Ch1 site in the inset of Fig. 6. In previous reports on the Se substitution in BiS$_2$-based compounds, the site selectivity of Se at the in-plane site was observed [24,32,33]. According to the relationship between in-plane disorder at the chalcogen site and superconductivity in BiS$_2$-based systems [16,24], we assume that Se substitution reduced the in-plane disorder at the S1 site and induced bulk superconductivity. Furthermore, the room-temperature Seebeck coefficient for the La$_2$O$_2$Bi$_3$Ag$_{0.6}$Sn$_{0.4}$S$_{5.7}$Se$_{0.3}$ sample was similar to those shown in Fig. 4 ($S$ = -25 μV/K). This also suggests that the bulk nature of superconductivity was induced by local structural optimization but not due to changes in carrier concentration.

Since we have reported the emergence of bulk superconductivity in $La_2O_2M_4S_6$-type layered oxychalcogenide $La_2O_2Bi_3Ag_{0.6}Sn_{0.4}S_{5.7}Se_{0.3}$, we can expect material development in related four-layer-type oxychalcogenide superconductors. Recently, a new superconductor $Bi_3O_2S_2Cl$ with the one-layer-type superconducting layer was discovered by Ruan et al. [34]. In Fig. 8, schematic images of typical one-layer-type (Fig. 8a), two-layer-type (Fig. 8b), and four-layer-type (Fig. 8c) are displayed for comparison. All the materials have the similar $RE_2O_2$ or $Bi_2O_2$ blocking layer. By changing the constituent elements in the superconducting layers, the thickness can be changed in this superconductor family. On the basis of these facts, we expect further development of materials with these crystal structures or novel materials with a superconducting layer with a different thickness (number of the superconducting layers in a unit cell).

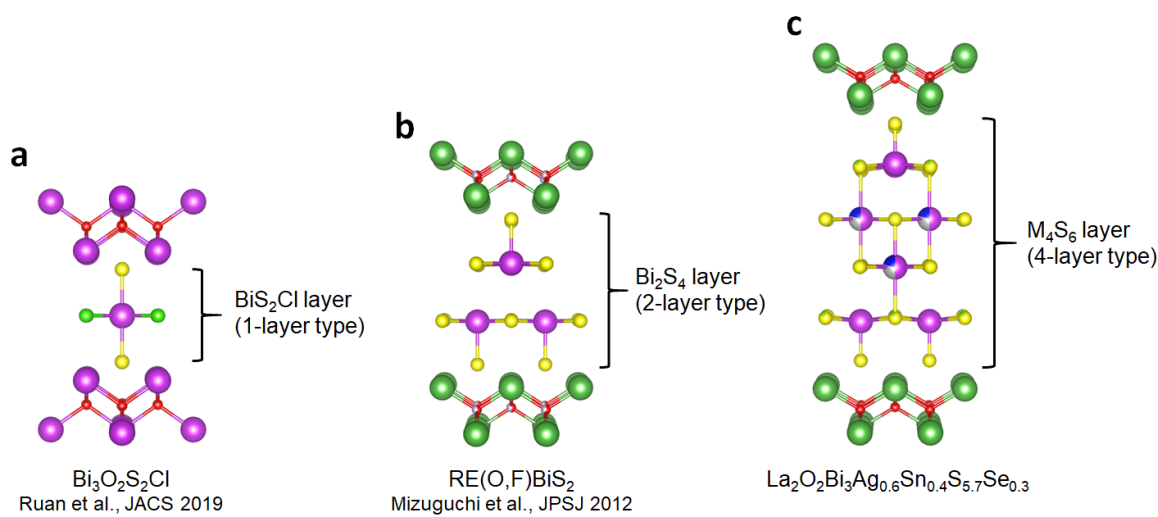

**Fig. 8. Comparison of crystal structure of typical layered oxychalcogenides with different thickness of superconducting layer.** **a.** Schematic image of the crystal structure of $Bi_3O_2S_2Cl$, which was recently discovered by Ruan et al. [34]. In this material, the conducting layer can be regarded as the one-layer-type $BiS_2Cl$. **b.** Schematic image of the crystal structure of the typical $BiS_2$-based superconductor $RE(O,F)BiS_2$. In the series, the conducting layer can be regarded as the two-layer-type $Bi_2S_4$ bilayer. **c.** Schematic image of the crystal structure of $La_2O_2Bi_3AgS_6$-type ($La_2O_2M_4S_6$-type) materials whose conducting layer can be regarded as the four-layer-type $M_4S_6$ layer.

**Methods**

      The polycrystalline samples of $La_2O_2Bi_3Ag_{1-x}Sn_xS_6$ with $x$ = 0, 0.1, 0.2, 0.3, 0.4, and 0.5 were prepared by a solid-state reaction method. The polycrystalline samples of Se-substituted $La_2O_2Bi_3Ag_{0.6}Sn_{0.4}S_{6-z}Se_z$ with $z$ = 0.3 and 0.6 were also prepared by a solid-state reaction method. Powders (or grains) of $Bi_2O_3$ (99.9%), $La_2S_3$ (99.9%), Sn (99.99%), and AgO (99.9%) and grains of Bi (99.999%), S (99.99%), and Se (99.99%) with a nominal composition of $La_2O_2Bi_3Ag_{1-x}Sn_xS_{6-z}Se_z$ were mixed in a pestle and mortar, pelletized, sealed in an evacuated quartz tube, and heated in an electric furnace. The heat treatment condition was 725 °C for 15 h for both samples. However, for $La_2O_2Bi_3Ag_{0.6}Sn_{0.4}S_{6-z}Se_z$, heating the sample to 725 °C in 1 h was needed to suppress the generation of impurity phases. The obtained samples were reground for homogeneity, pelletized, and heated in the same procedure. The phase purity of the prepared samples and the optimal annealing conditions were examined using X-ray diffraction (XRD) with a Cu-K$_\alpha$ radiation. The lattice parameters were determined using the Rietveld method with RIETAN-FP [35]. Schematic image of the crystal structure was drawn using VESTA [36]. The actual composition was analyzed by energy-dispersive X-ray spectroscopy (EDX) on scanning electron microscope TM3030 (Hitachi). The magnetic susceptibility measurements were carried out using a superconducting quantum interference device (SQUID) magnetometer (MPMS-3, Quantum Design). The susceptibility data were taken after both zero-field cooling (ZFC) and field cooling (FC). The temperature dependence of electrical resistivity [$\rho(T)$] was measured by four-terminal method on the Physical Property measurement system (PPMS, Quantum Design). The resistivity measurement down to 0.4 K was measured using a $^3$He probe platform of PPMS. The ADR system on PPMS was used for resistivity measurements down to 0.1 K. For clarity, we labeled the examined samples with the nominal compositions. The Seebeck coefficient was measured by a four-probe method on ZEM-3 (Advance RIKO) at 300 K.

**Author contributions**

R.J. synthesized the examined samples. R.J. and Y.M. evaluated the sample qualities and analyzed the crystal structure. R.J., T.D.M., Y.A., M.N., and I.T. measured electrical resistivity and analyzed the data. R.J. and Y.G. measured magnetic susceptibility and Seebeck coefficient. R.J. and Y.M. wrote the manuscript. Y.M. designed the research project.

**Data availability**

The datasets generated and analyzed during the current study are available from the corresponding author on reasonable request.

**Competing interests**

The authors declare no competing interests.

**Acknowledgments**

We gratefully appreciate O. Miura of Tokyo Metropolitan University for his technical support. This work was financially supported by grants in Aid for Scientific Research (KAKENHI) (Grant Nos. 15H05886, 15H05884, 16H04493, 17K19058, 16K05454, and 15H03693).